\documentclass[fleqn,twoside]{article}
\usepackage{espcrc2,epsfig}
\newcommand{\beq}{\begin{equation}}
\newcommand{\eeq}{\end{equation}}
\newcommand{\bea}{\begin{eqnarray}}
\newcommand{\eea}{\end{eqnarray}}
\newcommand{\mev}{\ {\rm MeV}}
\newcommand{\gev}{\ {\rm GeV}}
\newcommand{\ri}{{\rm RI/MOM}}
\newcommand{\msbar}{{\overline{\rm MS}}}
\newcommand{\ms}{m_s}
\newcommand{\ml}{m_\ell}
\newcommand{\Oa}{{\cal O}(a)}
\title{
Non-perturbative renormalization constants and light quark masses
\thanks{Presented by Vittorio Lubicz at ``Lattice 2002", Boston.}
\thanks{This work is supported in part by the European Network 
``Hadron Phenomenology and Lattice QCD", HPRN-CT-2000-00145.}}

\author{ {\tt SPQcdR} Collaboration\\
       D.~Becirevic\address[RM1]{Dip. di Fisica, Univ. di Roma ``La Sapienza"
       and INFN-Roma I, P.le A. Moro 2, I-00185 Rome, Italy.},
       V.~Gimenez\address[Val]{Dep. de F\'is.Te\`orica and IFIC, Univ. de 
       Val\`encia, Dr. Moliner 50, E-46100, Burjassot, Val\`encia, Spain.},
       V.~Lubicz\address[RM3]{Dip. di Fisica, Univ. di Roma Tre and 
       INFN-Roma III, Via della Vasca Navale 84, I-00146 Rome, Italy},
       G.~Martinelli\addressmark[RM1],
       M.~Papinutto\address{DESY, Theory Group, Notkestrasse 85, D-22607 
       Hamburg, Germany},
       J.~Reyes\addressmark[RM1] and
       C.~Tarantino\addressmark[RM3]}

\begin{document}

\begin{abstract}
We present the results of an extensive non-perturbative calculation of the 
renormalization constants of bilinear quark operators for the 
non-perturbatively $\Oa$-improved Wilson action. The results are obtained at 
four values of the lattice coupling, by using the RI/MOM and the Ward 
identities methods. A new non-perturbative renormalization technique, which is 
based on the study of the lattice correlation functions at short distance in 
$x$-space, is also numerically investigated. We then use our non-perturbative 
determination of the quark mass renormalization constants to compute the values 
of the strange and the average up/down quark masses. After performing an 
extrapolation to the continuum limit, we obtain $\ms^\msbar(2\gev) = (106\pm 2 
\pm 8)$~MeV and $\ml^\msbar(2\gev)=(4.4\pm 0.1\pm 0.4)$~MeV.
\vspace{-0.5cm}
\end{abstract}

\maketitle

Although the calculation of renormalization constants (RCs) can be performed 
in perturbation theory, the implementation of non-perturbative renormalization 
(NPR) techniques has become an essential ingredient of lattice QCD 
calculations. These techniques allow to control the systematic error associated 
with the determination of the RCs at the level of accuracy reached at present 
by the statistical and other systematic uncertainties in lattice calculations.

In this talk we present the results of an extensive NP 
calculation~\cite{zetapap} of the RCs $Z_V$, $Z_A$, $Z_S$, $Z_P$ and $Z_T$ of 
the bilinear quark operators, for the non-perturbatively $\Oa$-improved 
Wilson action~\cite{alpha}. These results are obtained at four values of the 
lattice coupling, in the range $6.0\le \beta \le 6.45$, by using the $\ri$ 
method~\cite{rimom} and, for the scale independent RCs, the Ward identity (WI) 
approach~\cite{boch}. A NP $\ri$ determination of the RCs of the $\Delta F=2$ 
four-fermion operators has been also presented by J.~Reyes at this 
Conference~\cite{reyes}.

The quark mass renormalization constants computed in this study have been used 
to calculate the strange and the average up/down quark masses~\cite{qm_spqr}. 
That calculation, which is completely $\Oa$-improved, uses NPR and involves an 
extrapolation to the continuum limit, represents at present one of the most 
accurate determination of the light quark masses within the quenched 
approximation.

Finally, we present the results of a preliminary numerical investigation of a 
NPR technique based on the study of the lattice correlation functions at 
short distance in $x$-space (XS)~\cite{XS}.

\vspace*{-0.2cm}
\section{RENORMALIZATION CONSTANTS}
The lattice parameters used in this study are summarized in
table~\ref{table:details}.
The values of the inverse lattice spacing given in the table have been 
determined from the study of the static quark anti-quark 
potential~\cite{necco}, fixing in input $a^{-1}(\beta=6)=2.0(1)\gev$.
\begin{table}[t]
\caption{\small \sl Summary of the lattice parameters used in this work.}
\label{table:details}
\begin{tabular}{|ccccc|}  \hline \hline
$ \beta $ & $a^{-1}$(GeV) & $ L^3 \times T $ & 
$ \#\ \kappa_\ell$ & $ \#\ {\rm Confs}$ \\ \hline 
6.0 & 2.00(10) & $16^3 \times 52$ & 4 & 500 \\
6.0 & 2.00(10) & $24^3 \times 64$ & 3 & 340 \\
6.2 & 2.75(14) & $24^3 \times 64$ & 4 & 200 \\
6.4 & 3.63(18) & $32^3 \times 70$ & 4 & 150 \\
6.45& 3.87(19) & $32^3 \times 70$ & 4 & 100 \\
\hline \hline
\end{tabular}
\vspace*{-0.8cm}
\end{table}

In order to study finite size effects, two independent simulations on different 
lattice sizes have been performed at $\beta=6.0$. The smallest size corresponds 
approximately to the same physical volume used at the larger values of $\beta$. 
From this analysis, we find that finite volume effects in the calculation are 
limited in the range between 1\% ($Z_V$, $Z_A$) and 3\% ($Z_S$, $Z_P$, $Z_T$).

The $\ri$ determination of the RCs is based on the study of the lattice 
correlation functions at large $p^2$ and in the chiral limit. The
renormalization condition, in the $\ri$ scheme, ensures that leading 
$\Oa$-discretization effects vanish in these limits~\cite{zetapap}. Therefore, 
the improvement of the external quark fields and of the composite operators is 
not necessary for this determination. The on-shell improvement of the vector 
and axial-vector current operators is required, instead, when the RCs are 
computed from the study of the lattice WIs. In this case, we use the 
coefficients $c_V$ and $c_A$ determined non-perturbatively in 
refs.~\cite{alpha,lanl}.

\begin{figure}[t]
\hspace*{-0.6cm}
\begin{tabular}{cc}
\epsfxsize4.0cm\epsffile{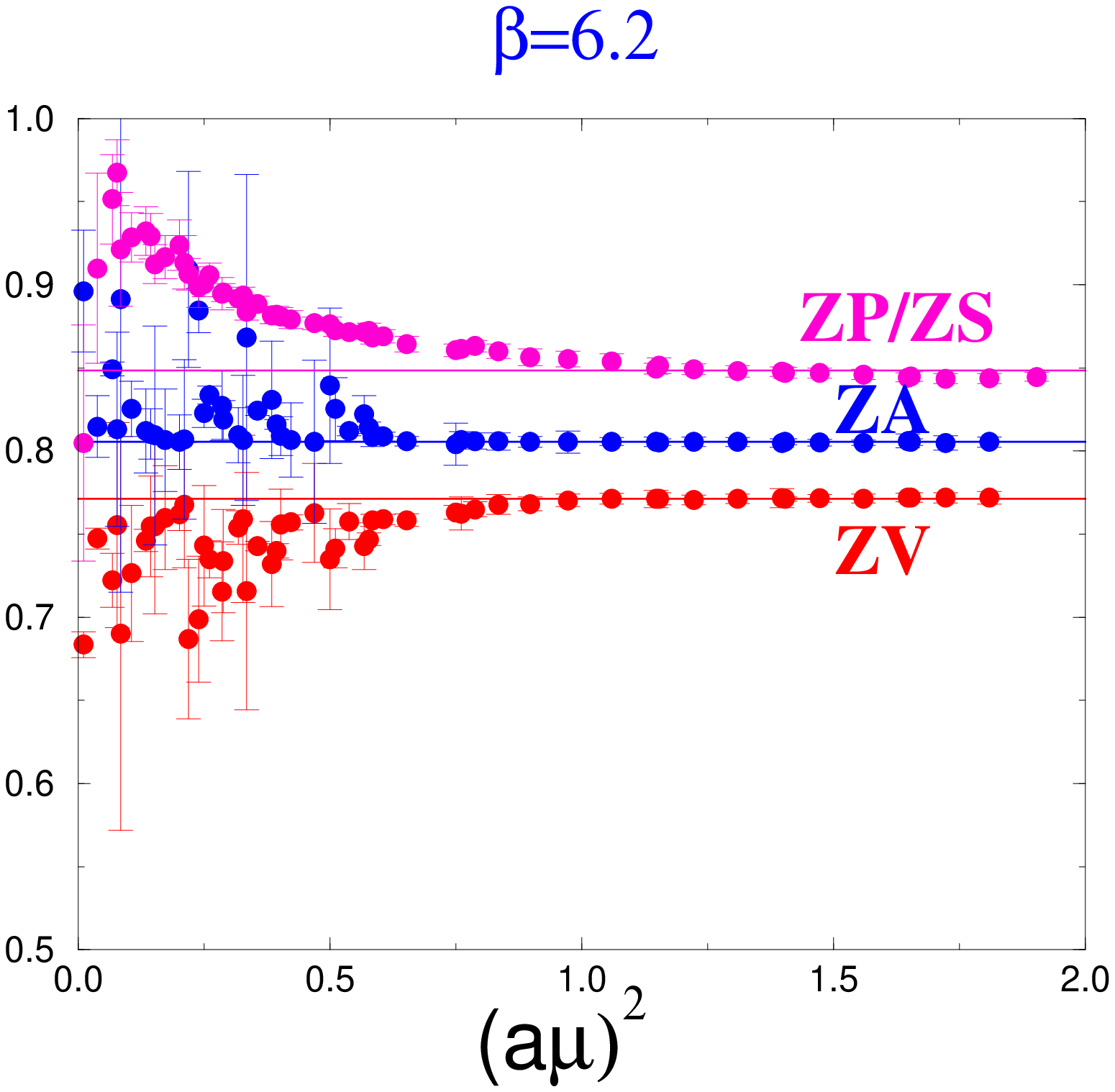} &
\hspace*{-0.5cm}
\epsfxsize4.0cm\epsffile{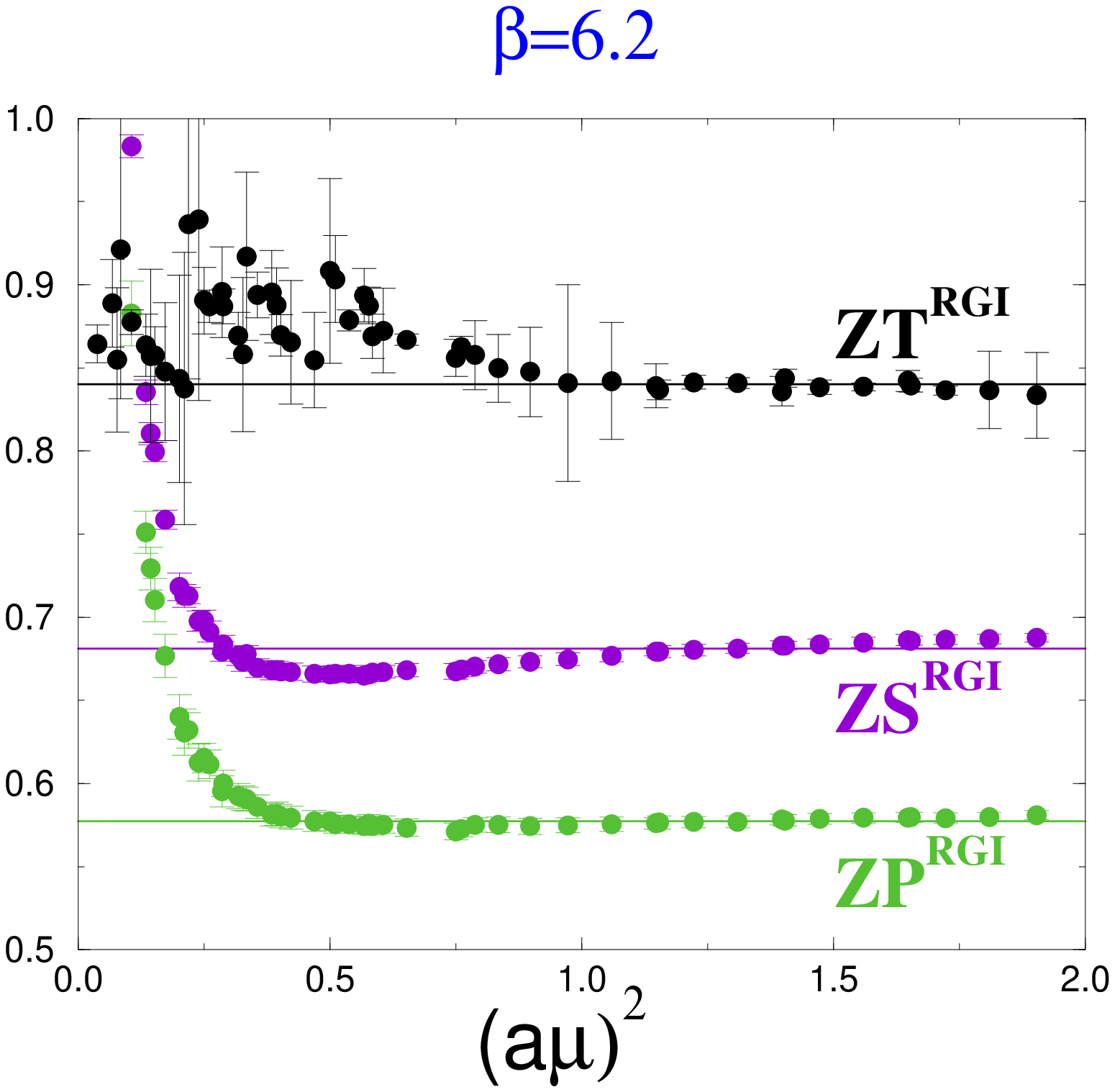} \\
\end{tabular}
\vspace*{-1.2cm}
\caption{\label{fig:zall62}{\small \sl The scale independent RCs $Z_V$, $Z_A$ 
and the ratio $Z_P/Z_S$ (left) and the RGI combinations $Z_P^{RGI}$, 
$Z_S^{RGI}$ and $Z_T^{RGI}$ (right) as a function of the renormalization scale, 
at $\beta=6.2$. The solid lines represent the results of a constant fit to the 
data.}}
\vspace*{-0.8cm}
\end{figure}

The results for the scale independent RCs $Z_V$ and $Z_A$, and for the ratio 
$Z_P/Z_S$, obtained with the $\ri$ method from the simulation at $\beta=6.2$ 
are shown, as an example, in fig.\ref{fig:zall62} (left) as a function of the 
renormalization scale. 
The good quality of the plateau indicates that ${\cal O}(a^2)$ discretization 
effects are well under control. In fig.\ref{fig:zall62} (right) we also show 
the renormalization group invariant (RGI) combinations $Z_{\cal O}^{RGI} = 
Z_{\cal O}(\mu)/C_{\cal O}(\mu)$, for ${\cal O}=S,P,T$, where the evolution 
function $C_{\cal O}(\mu)$ is introduced to cancel the scale dependence of the 
RCs. In the $\ri$ scheme, these functions are known at the NLO in perturbation 
theory for $Z_T$ and at the N$^3$LO for $Z_S$ and $Z_P$~\cite{chetyr}. From the 
quality of the plateau, discretization effects appear to be very small, even at 
quite large values of $(a\mu)^2$. Our estimates of the RCs have been 
obtained by fitting to a constant the results shown in fig.\ref{fig:zall62} 
(and similarly for the other $\beta$s). The uncertainty due to the residual 
scale dependence observed in the plots has been used to evaluate the systematic 
errors.

\begin{figure}[t]
\begin{center}
\begin{tabular}{c}
\epsfxsize4.0cm
\epsffile{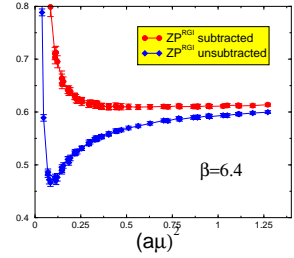} \\
\end{tabular}
\end{center}
\vspace*{-1.2cm}
\caption{\label{fig:zpnosub}{\small \sl $Z_P^{RGI}$ as obtained with and 
without the subtraction of the Goldstone pole, at $\beta=6.4$.}}
\vspace*{-0.8cm}
\end{figure}
The $\ri$ approach relies on the fact that NP contributions to the Green 
functions vanish asymptotically at large $p^2$. Special care, however, must be 
taken in the study of the pseudoscalar Green function $\Gamma_P$ since in this 
case, due to the coupling to the Goldstone boson, the leading power 
suppressed contribution is divergent in the chiral limit~\cite{rimom,alain}. In 
order to improve the convergence of $\Gamma_P$ to its asymptotic behaviour at 
large $p^2$, we have subtracted the Goldstone pole contribution by constructing 
the combinations~\cite{gv}
\beq
\Gamma_P^{\rm SUB}(m_1,m_2) = \frac{m_1 \Gamma_P(m_1) - m_2 \Gamma_P(m_2)}
{m_1-m_2} \,,
\eeq
where $m_1$ and $m_2$ are different quark masses. The effect of this 
subtraction on $Z_P^{RGI}$ is shown in fig.\ref{fig:zpnosub}. Though both the 
unsubtracted and the subtracted determinations of $Z_P^{RGI}$ converge to the 
same value at large $p^2$, a clear plateau is never reached in the first case, 
in the region of momenta explored in this study.
\begin{table*}[t]
\caption{\small \sl Values of the RCs obtained with the $\ri$ method. The 
scheme and scale dependent constants, $Z_S$, $Z_P$ and $Z_T$ are given in the 
$\ri$ scheme at the scale $\mu=1/a$.}
\label{table:rc_rimom}
\begin{tabular}{||c|cccc||c|cccc||}
\hline
$\beta$ & 6.0 & 6.2 & 6.4 & 6.45 &
$\beta$ & 6.0 & 6.2 & 6.4 & 6.45 \\
\hline\hline
$Z_V$ & 0.766(2) & 0.775(3) & 0.795(2) & 0.797(3) &
$Z_S$ & 0.668(13)& 0.677(9) & 0.696(6) & 0.707(8) \\
\hline
$Z_A$ & 0.804(2) & 0.809(3) & 0.824(2) & 0.825(4) &
$Z_P$ & 0.535(12)& 0.564(4) & 0.600(5) & 0.612(9) \\
\hline
$Z_P/Z_S$ & 0.804(6) & 0.831(3) & 0.862(3) & 0.867(3) &
$Z_T$     & 0.833(2) & 0.847(2) & 0.867(6) & 0.867(5) \\
\hline
\end{tabular}
\vspace*{-0.3cm}
\end{table*}
\begin{table}[t]
\caption{\small \sl Values of the RCs obtained with the WI method.}
\label{table:rc_wi}
\begin{tabular}{||c|ccc||}
\hline
$\beta$ &  6.0 & 6.2 & 6.4 \\
\hline
$Z_V$     & 0.774(4)  & 0.789(2) & 0.804(2)  \\ 
$Z_A$     & 0.841(17) & 0.811(5) & 0.843(10) \\
$Z_P/Z_S$ & 0.845(20) & 0.863(5) & 0.911(10) \\
\hline
\end{tabular}
\vspace*{-0.5cm}
\end{table}

\begin{figure}[t]
\hspace*{-0.6cm}
\begin{tabular}{cc}
\epsfxsize4.0cm\epsffile{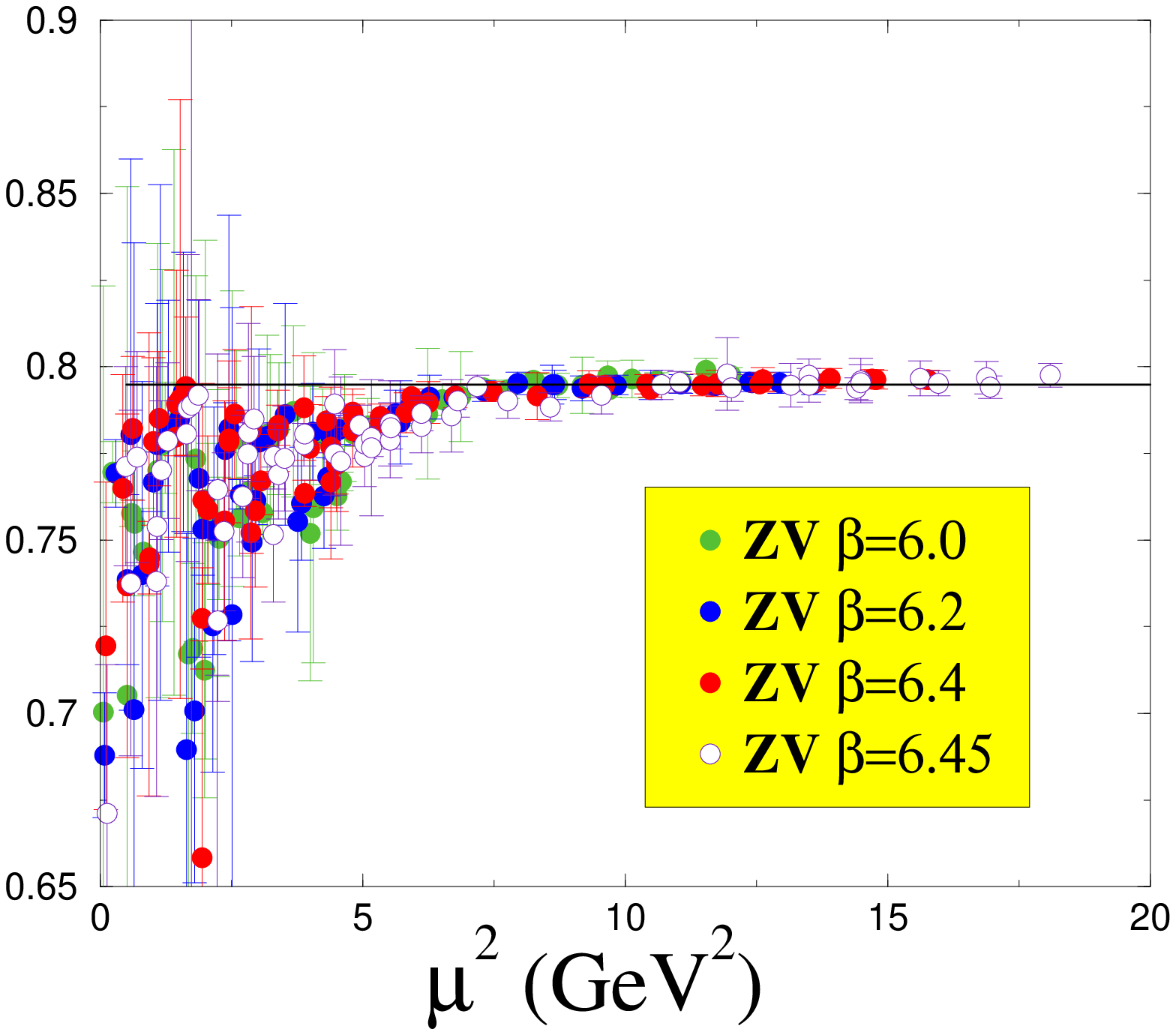} &
\hspace*{-0.5cm}
\epsfxsize4.0cm\epsffile{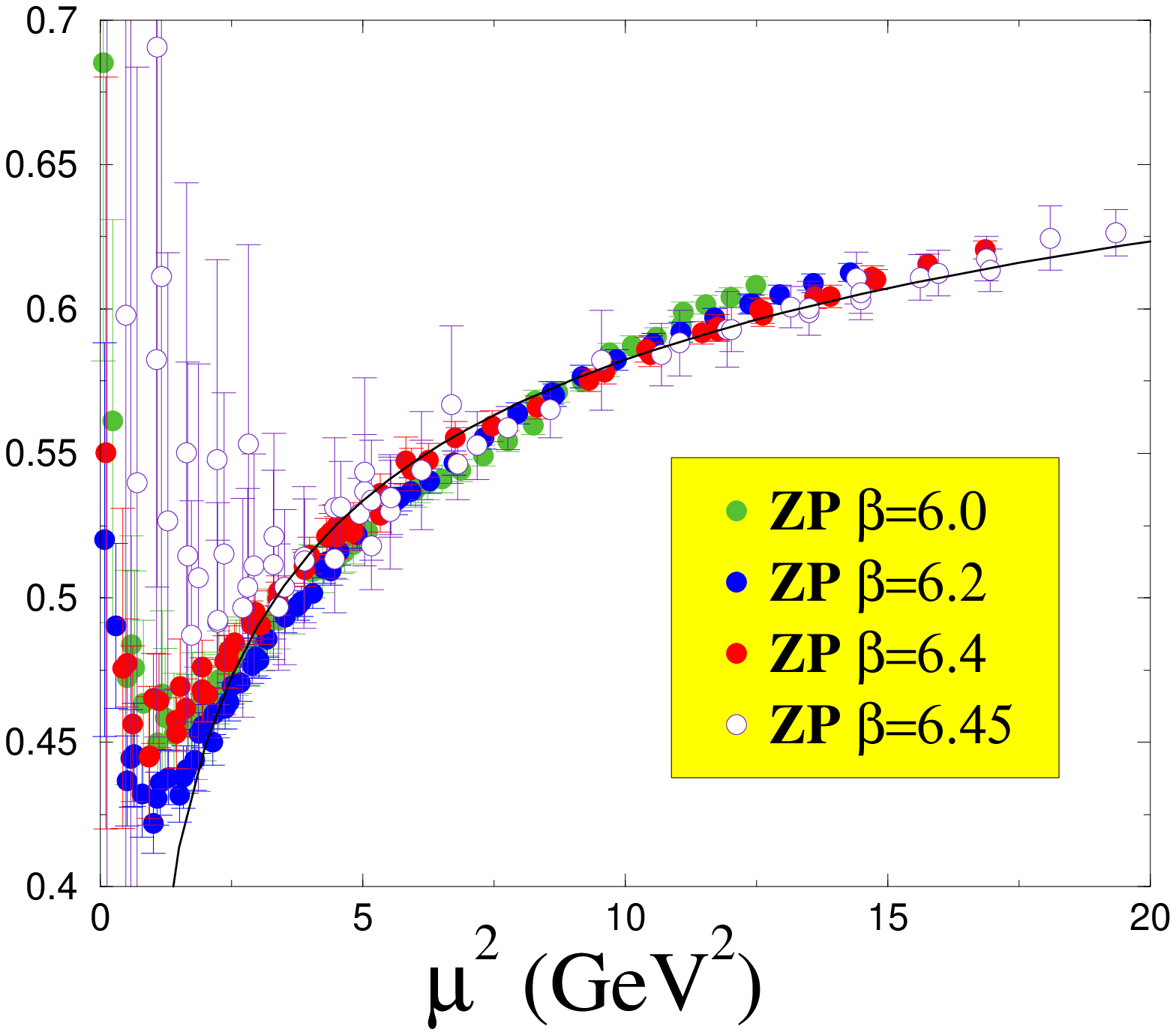} \\
\end{tabular}
\vspace*{-1.2cm}
\caption{\label{fig:zrisc}{\small \sl $Z_V$ and $Z_P$ at the four values of the 
lattice coupling as a function of the renormalization scale. The results have 
been rescaled by the factor $R(a,a')$ defined in eq.~(\ref{eq:erre}). The solid 
lines represent the scale dependence predicted by the corresponding anomalous 
dimensions.}}
\vspace*{-0.8cm}
\end{figure}

Discretization effects can also be investigated by comparing the results for 
the RCs obtained at different values of the lattice spacing. Though the RCs
depend on the lattice spacing (the UV cutoff in the lattice regularization), 
their dependence on the renormalization scale is the same for all $a$, and it 
is only fixed by the anomalous dimension of the relevant operators. Therefore, 
the scale dependence should cancel in the ratio
\beq
\label{eq:erre}
Z(a,\mu)/Z(a',\mu) = R(a,a')\, ,
\eeq
up to discretization effects. In fig.\ref{fig:zrisc} we show as an example the 
results for the RCs $Z_V$ and $Z_P$ obtained at four values of $\beta$ as 
a function of the renormalization scale. Each RC, $Z(a',\mu)$, has been 
rescaled by the factor $R(a,a')$ of eq.~(\ref{eq:erre}), where we have chosen 
as a reference scale the value of the lattice spacing at $\beta=6.4$. In 
fig.\ref{fig:zrisc} we observe that the results obtained at the different 
values of $\beta$ all lie on the same universal curve, thus confirming that 
discretization effects are reasonably small, within the statistical errors. 
The figure also shows that the renormalization scale dependence is in very good 
agreement with the one predicted by the anomalous dimensions of the relevant 
operators.

\begin{figure}[t]
\hspace*{-0.6cm}
\begin{tabular}{cc}
\epsfxsize4.0cm\epsffile{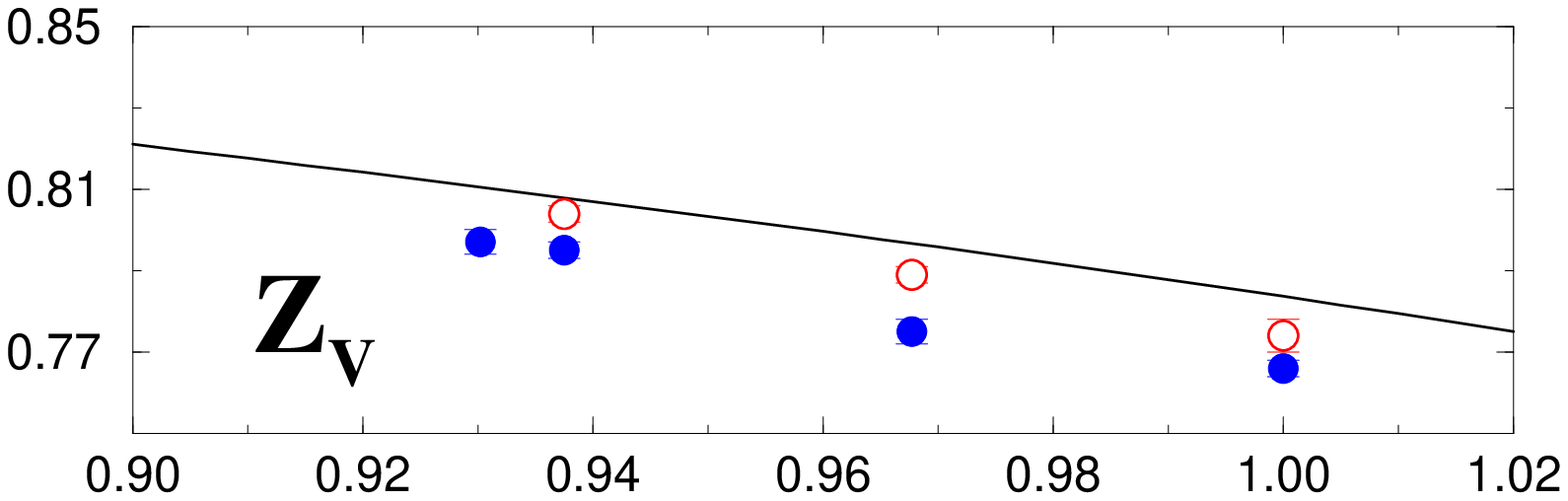} &
\hspace*{-0.5cm}
\epsfxsize4.0cm\epsffile{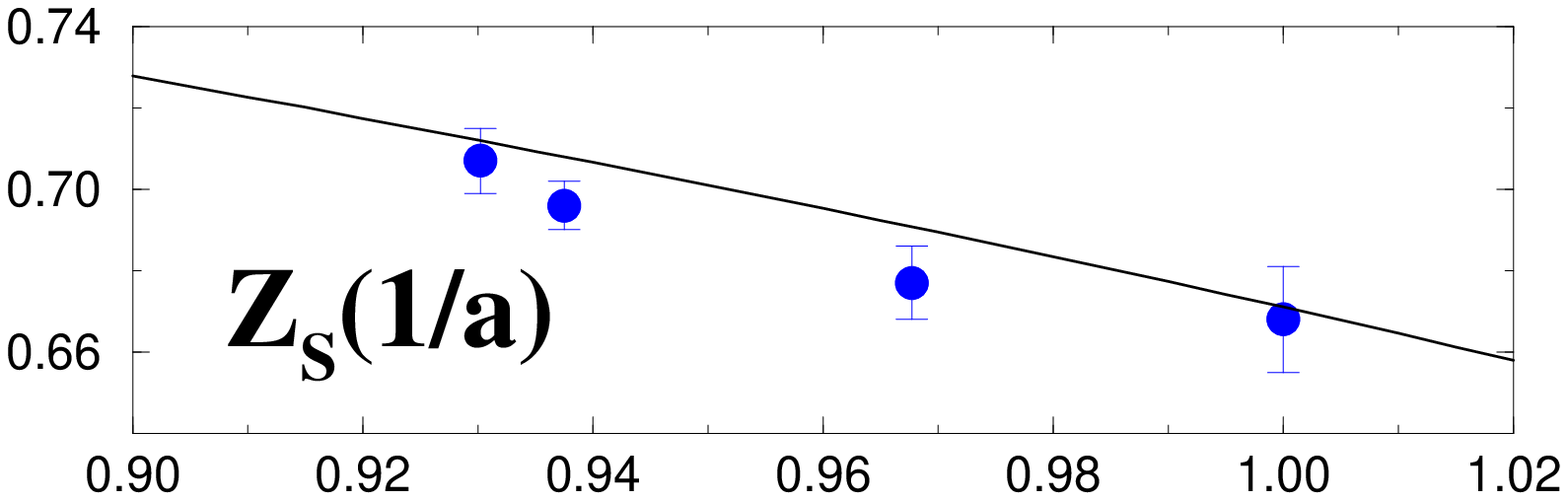} \\
\epsfxsize4.0cm\epsffile{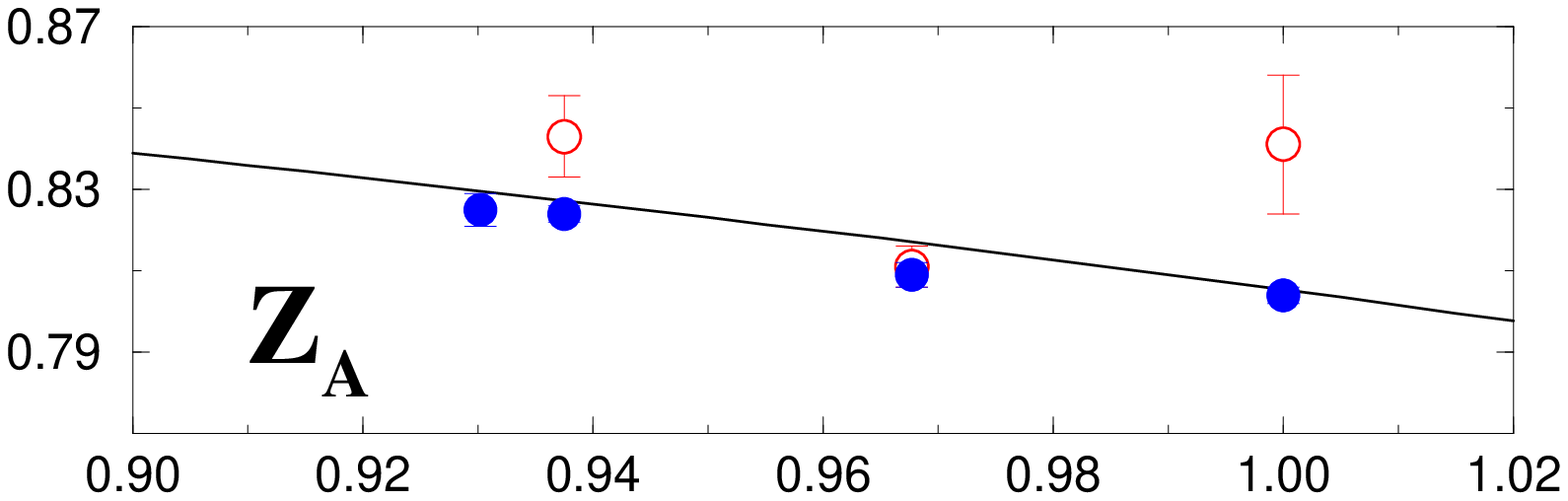} &
\hspace*{-0.5cm}
\epsfxsize4.0cm\epsffile{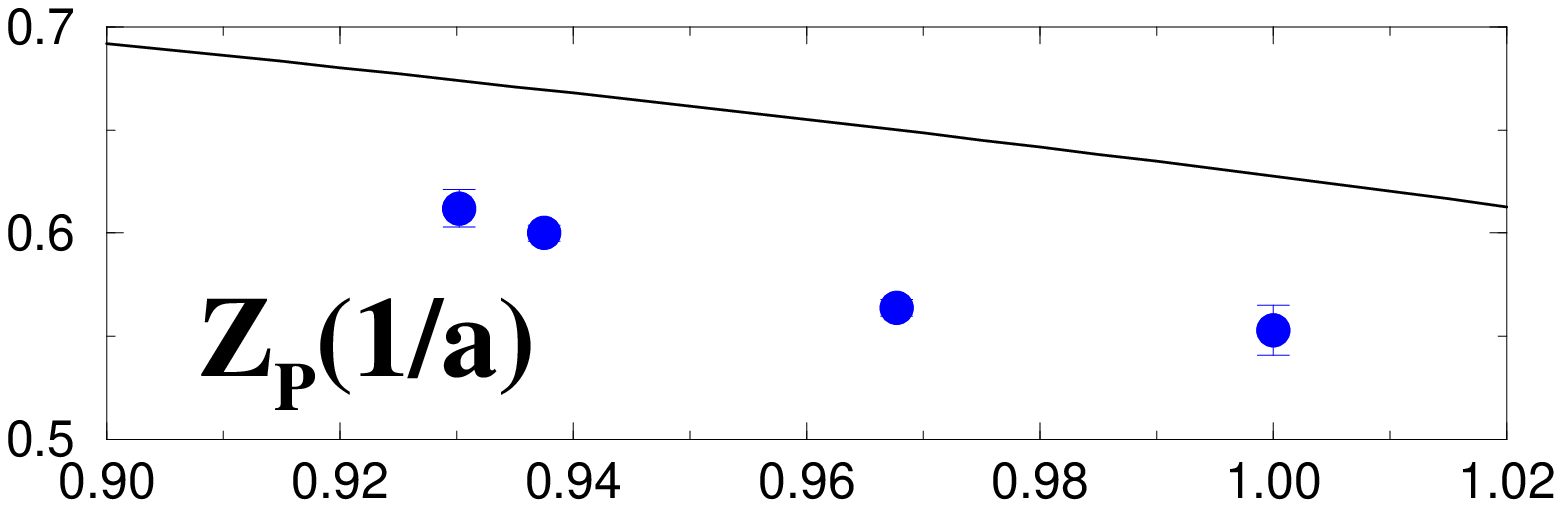} \\
\epsfxsize4.0cm\epsffile{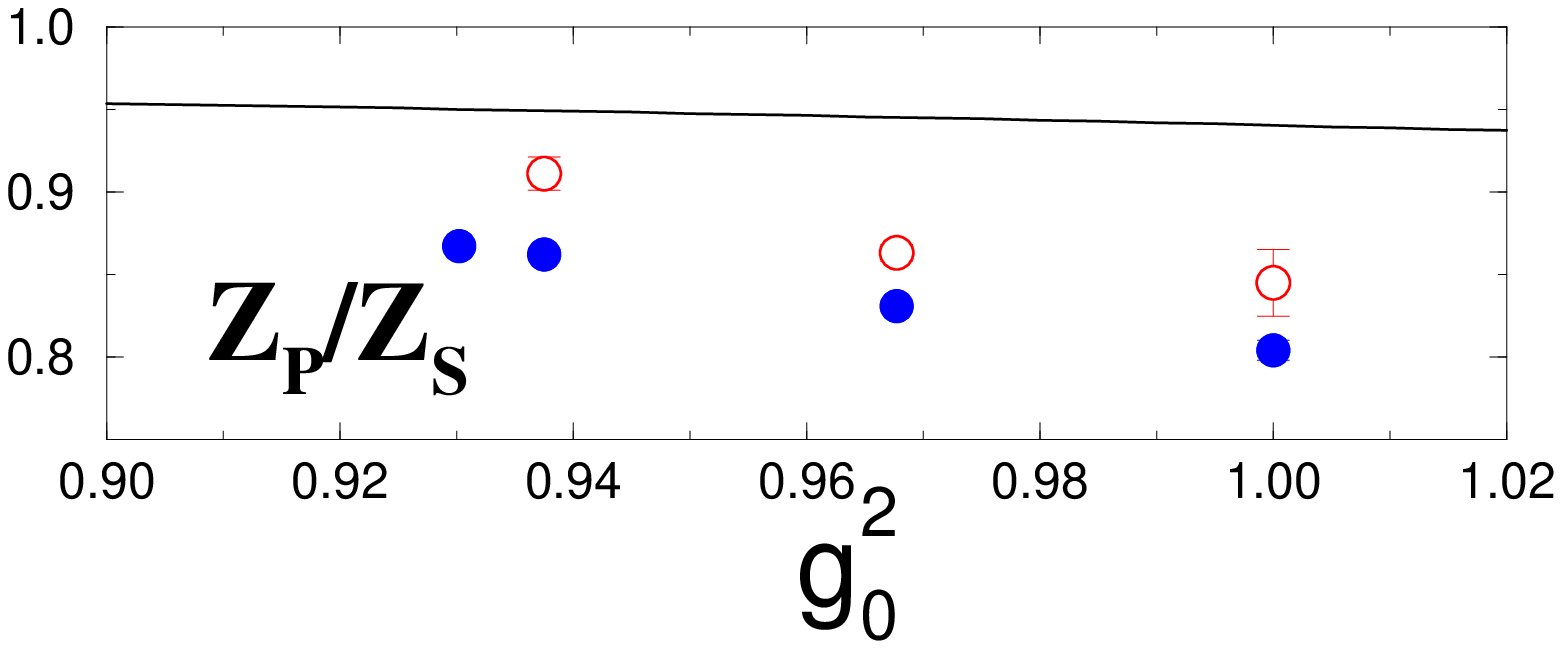} &
\hspace*{-0.5cm}
\epsfxsize4.0cm\epsffile{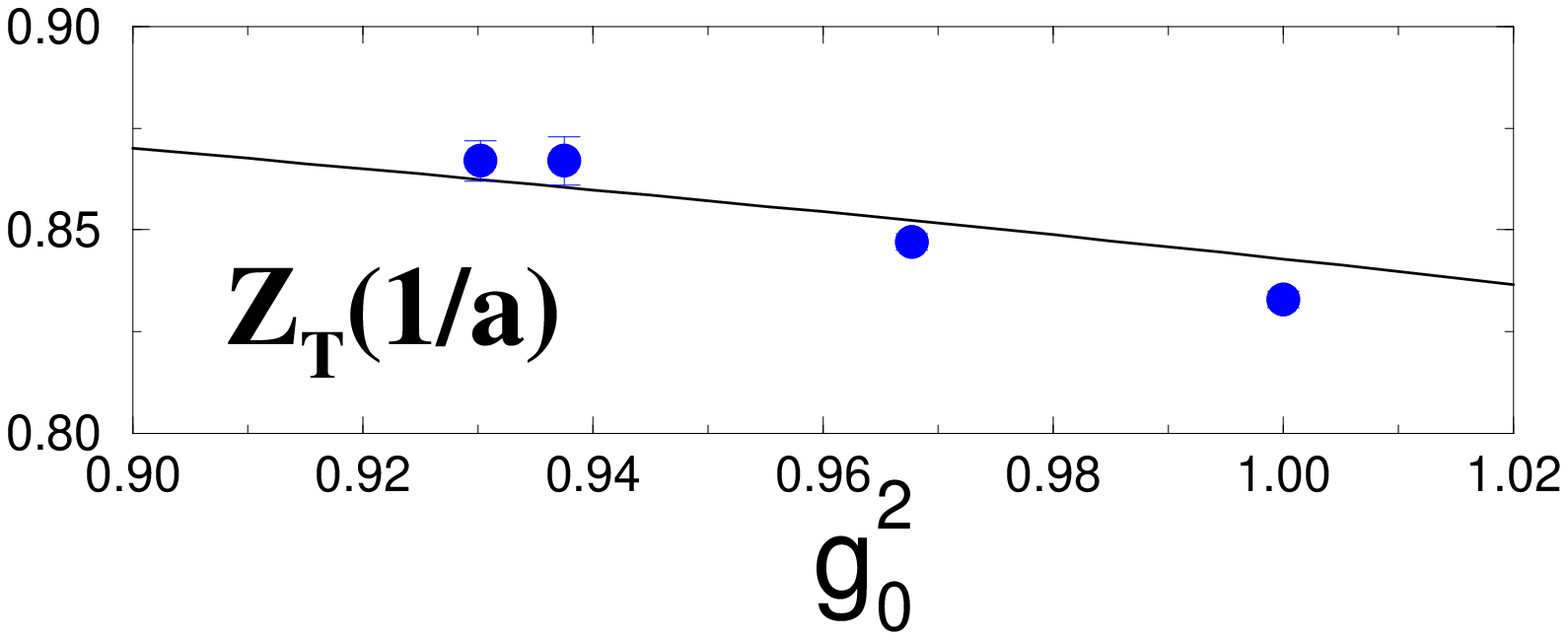} \\
\end{tabular}
\vspace*{-1.2cm}
\caption{\label{fig:zeta}{\small \sl Values of the RCs obtained from the $\ri$ 
method (filled circles), the WI method (empty circles) and 1-loop BPT (solid 
lines).}}
\vspace*{-0.8cm}
\end{figure}

Our results for the RCs of the bilinear quark operators determined with the
$\ri$ and the WI methods are collected in tables~\ref{table:rc_rimom} and 
\ref{table:rc_wi} respectively. These results are shown in fig.\ref{fig:zeta}, 
where they are also compared with the predictions of 1-loop boosted 
perturbation theory (BPT).
We observe a good agreement between the $\ri$ and the WI determinations for the 
cases in which the latter method can be used, namely $Z_V$, $Z_A$ and $Z_P/Z_S$.
There is also a general good agreement between NP results and BPT. A notable 
exception is the RC of the pseudoscalar density $Z_P$, whose NP determinations 
differ from the perturbative prediction by approximately 10-15\%.

\vspace*{-0.2cm}
\section{LIGHT QUARK MASSES}
An important application of the NPR is the lattice determination of the quark 
masses. We have computed the strange and the average up/down quark masses by
using the standard vector and axial-vector WI methods, for which the relevant 
RCs are $1/Z_S$ and $Z_A/Z_P$ respectively. We have used the same sets of gauge 
configurations and quark propagators used for the calculation of the RCs. The 
physical values of the light quark masses have been fixed by using the 
experimental pion and kaon masses. Particular attention, in the calculation, 
has been dedicated to control and reduce the systematic uncertainties, within 
the quenched approximation. The final results have been obtained by performing
an extrapolation to the continuum limit. All details of this calculation are 
discussed in ref.~\cite{qm_spqr}. Here we only quote our final results, which 
read:
\bea
&& \ms^\msbar (2\gev) = (106\pm 2 \pm 8)\mev \nonumber \\
&& \ml^\msbar (2\gev) = (4.4\pm 0.1 \pm 0.4)\mev \,.
\eea

\vspace*{-0.2cm}
\section{THE X-SPACE METHOD}
In the XS method~\cite{XS} one imposes the following renormalization condition
\beq
\label{xs}
Z_O(x_0/a) \, \langle O(x)O(0)\rangle \vert_{x^2=x_0^2} = 
\langle O(x_0)O(0)\rangle_{\rm cont}
\eeq
where $x_0$ is a short distance scale. The symbol $\langle \ldots \rangle_{\rm 
cont}$ denotes the Green function renormalized in a continuum scheme (the 
simplest choice is to fix this function to its value in the free theory).

With respect to the $\ri$ method, the XS approach presents an important 
theoretical advantage: the correlation functions in eq.~(\ref{xs}) are i) gauge 
invariant and ii) not affected by the presence of contact terms for $x_0\neq 0$.
This allows to neglect, in the renormalization procedure, the mixing with non 
gauge invariant operators and/or operators vanishing by the equation of motion.

\begin{figure}[t]
\hspace*{-0.6cm}
\begin{tabular}{cc}
\epsfxsize4.0cm\epsffile{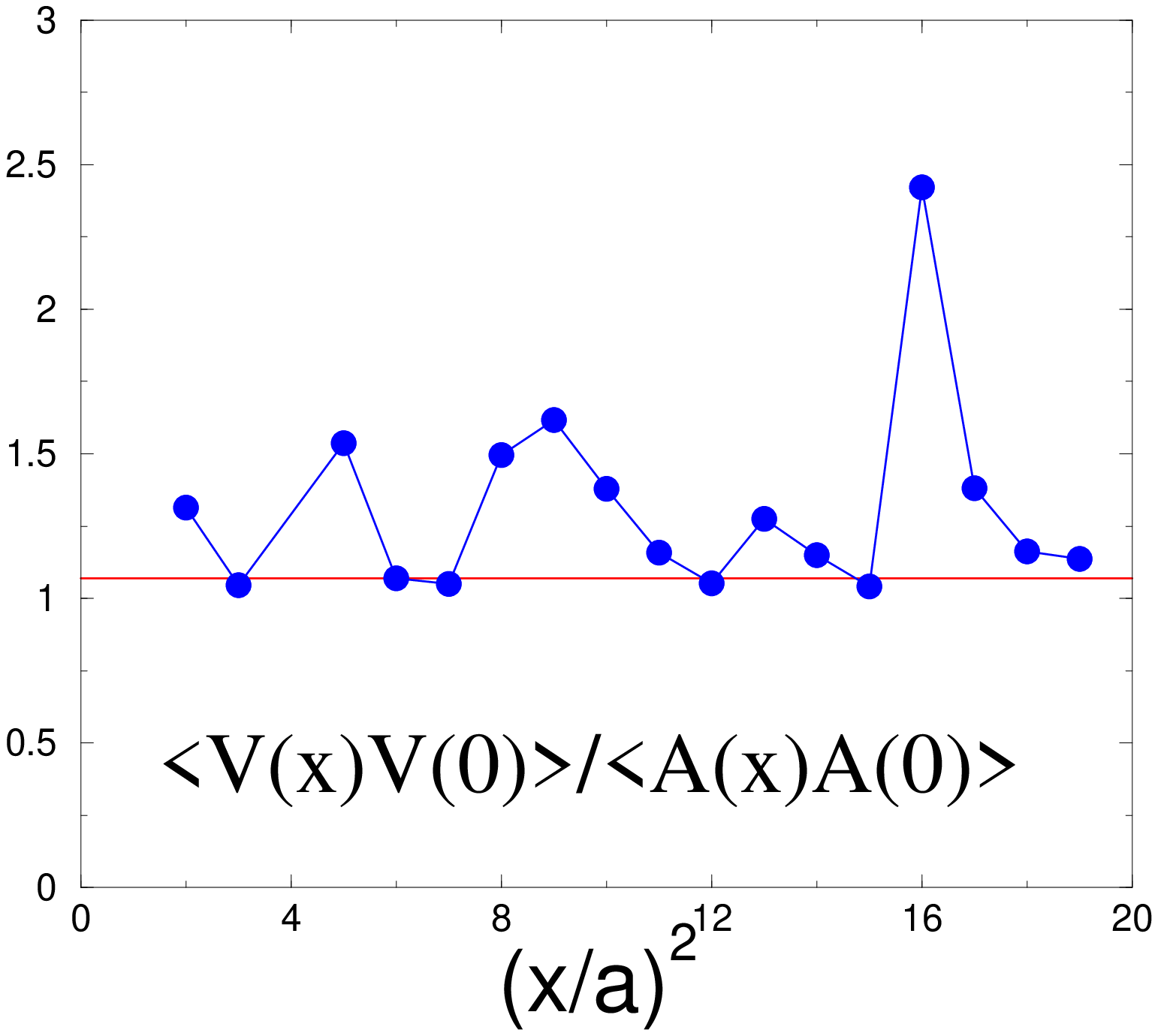} &
\hspace*{-0.5cm}
\epsfxsize4.0cm\epsffile{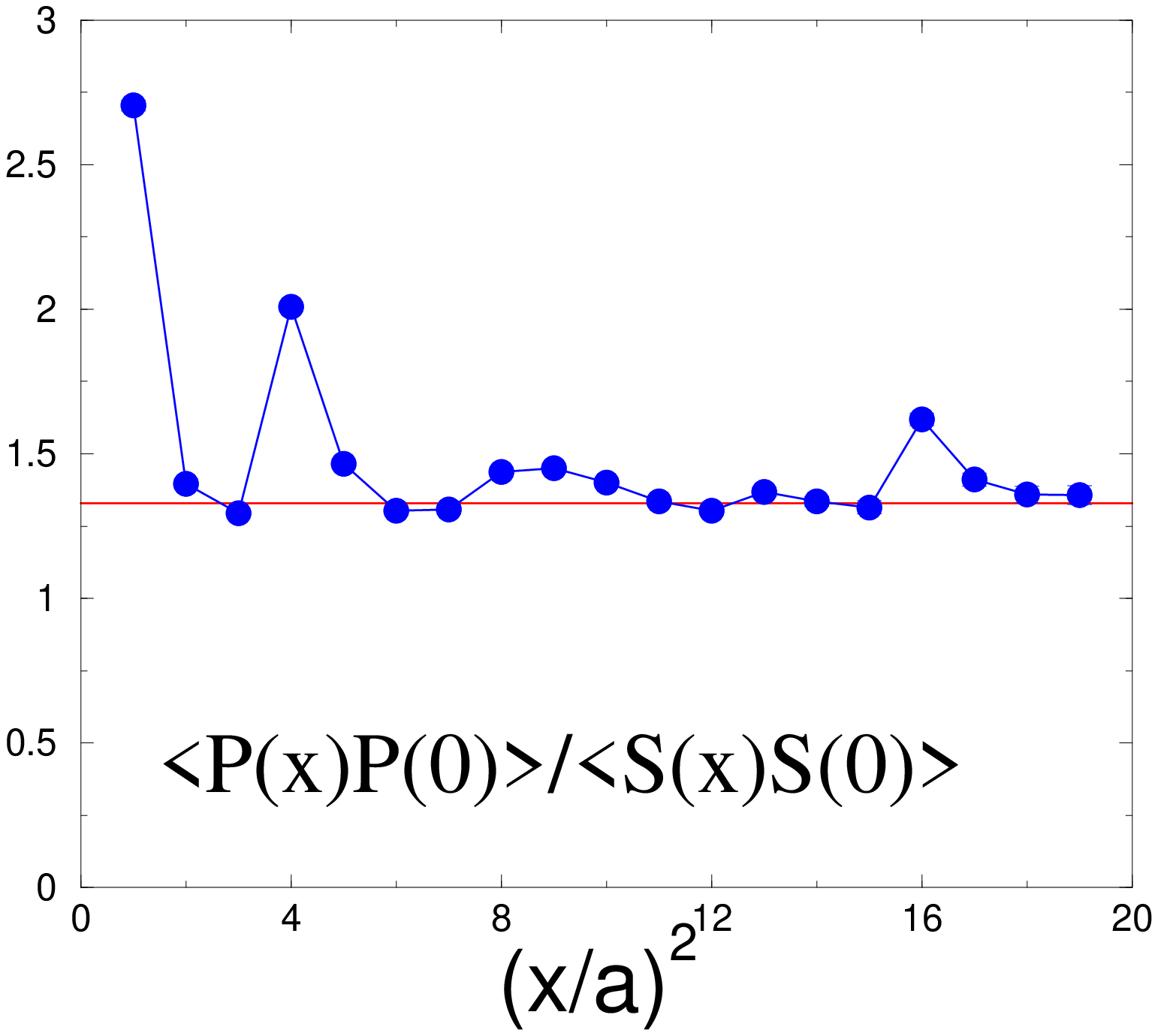} \\
\end{tabular}
\vspace*{-1.2cm}
\caption{\label{fig:spazioX}{\small \sl Ratios of 2-point correlation functions
as a function of the distance $(x/a)^2$. For comparison, the $\ri$ estimates of 
$(Z_A/Z_V)^2$ and $(Z_S/Z_P)^2$ are also shown with solid lines.}}
\vspace*{-0.8cm}
\end{figure}

As a preliminary study of the XS method we show in fig.\ref{fig:spazioX} the
results for the ratios $\langle V(x)V(0)\rangle/\langle A(x)A(0)\rangle$ and 
$\langle P(x)P(0)\rangle$/ $\langle S(x)S(0)\rangle$, as a function of the 
distance $x^2$ (points corresponding to the same value of $x^2$ have been 
averaged together). These quantities provide directly the scheme and scale 
independent ratios $(Z_A/Z_V)^2$ and $(Z_S/Z_P)^2$. For comparison, the $\ri$ 
estimates of this RCs are also shown in the figure. We note that, though the 
results obtained with the XS method are consistent with the expectations, 
they are affected by large systematic uncertainties. Understanding these 
uncertainties requires further investigations.

\end{document}